\documentclass[11pt,superscriptaddress,aps,prd,preprint,showpacs]{revtex4}

\usepackage[dvips]{graphicx}
\usepackage{amsfonts}
\usepackage{slashed}
\usepackage[utf8x]{inputenc}
\usepackage{amsmath}
\usepackage{hyperref}

\newcommand{\¡}{\negthickspace} 

\begin{document}

\title{Horava-Lifshitz four-fermion model revisited and dynamical symmetry breaking}

\author{T. Mariz}
\affiliation{Instituto de F\'{\i}sica, Universidade Federal de Alagoas,\\ 57072-900, Macei\'o, Alagoas, Brazil}
\email{tmariz@fis.ufal.br}

\author{J. R. Nascimento}
\affiliation{Departamento de F\'{\i}sica, Universidade Federal da Para\'{\i}ba,\\
 Caixa Postal 5008, 58051-970, Jo\~ao Pessoa, Para\'{\i}ba, Brazil}
\email{jroberto,petrov@fisica.ufpb.br}

\author{A. Yu. Petrov}
\affiliation{Departamento de F\'{\i}sica, Universidade Federal da Para\'{\i}ba,\\
 Caixa Postal 5008, 58051-970, Jo\~ao Pessoa, Para\'{\i}ba, Brazil}
\email{jroberto,petrov@fisica.ufpb.br}

\begin{abstract}
In this paper, we develop studies of the dynamical symmetry breaking in the Horava-Lifshitz four-fermion model for the specific case $z=3$ and explicitly demonstrate that for various space-time dimensions, one could arrive at the theory displaying both dynamical generation of the Lorentz symmetry for the kinetic term and arising the positively defined potential at the same time. At the same time, for $D=3$, the Lorentz invariant Chern-Simons term is generated.
\end{abstract}


\maketitle

\section{Introduction}

Dynamical symmetry breaking, i.e., spontaneous breaking of a continuous symmetry generated by quantum corrections, is an interesting phenomenon occurring in various field theory models, allowing for many fundamental effects such as mass generation (for reviews and  phenomenological applications, see \cite{Poggio,Chuk}). The paradigmatic example of a theory displaying such behavior is the Gross-Neveu model \cite{GN}. Dynamical symmetry breaking played a crucial role within the formulation of a great unified theory \cite{Wein,Suss}. All this justifies the interest and importance of studying the dynamical symmetry breaking within various contexts. Among important results, one can mention the dynamical breaking of gauge symmetry \cite{Appel}, supersymmetry \cite{Intri}, and Lorentz symmetry \cite{Seif}. One of the interesting applications of this methodology includes studies of theories with space-time anisotropy, also known as Horava-Lifshitz-like theories \cite{HL}, interest to which strongly increased in recent years (various studies of these theories are presented in \cite{difpapers}, see also references therein). In these theories, the role of the continuous symmetry is played by rotational symmetry $O(d)$, for a $d$-dimensional space.

In our previous paper \cite{ourprev}, the dynamical symmetry breaking has been considered in a $z=2n+1$ four-fermion Horava-Lifshitz theory. It has been shown that the effective potential generated in the one-loop approximation possesses a set of minima allowing for spontaneous breaking of rotational symmetry. However, the scheme proposed in that paper needs some improvements. Indeed, it was claimed in \cite{ourprev} that, starting from the four-fermion Lagrangian proposed, one can arrive at two possible situations. Within the first case, under a specific gauge  condition $(\partial_i A_0)^2=0$, arising of the Maxwell term and hence dynamical restoring of the Lorentz symmetry in the low-energy limit takes place, which is reasonable within the concept of emergent dynamics \cite{Bjorken}, but the potential of the vector field is not positive definite and hence does not display set of minima necessary for dynamical symmetry breaking. Within the second case, one arrives at arising a positively defined potential possessing a set of minima, whereas the Lorentz invariant Maxwell-like kinetic term, whose arising could be very natural to provide a consistent low-energy effective behavior, cannot be generated. Clearly, none of these situations can be treated as a completely satisfactory from the physical viewpoint.

Hence, a natural question is -- whether it is still possible to have a situation where both Lorentz symmetric Maxwell kinetic term and the positively defined potential with a continuous set of minima are generated at the same time. In this paper, we discuss such a possibility and find a physically consistent situation where one can conciliate arising of a potential allowing for spontaneous breaking of rotational symmetry with arising the usual Maxwell kinetic term for the vector field.

The structure of the paper looks like follows. In section 2, we introduce the action of the $z=3$ Lifshitz four-fermion model, write down the generating functional and the interaction vertices. In section 3, we calculate the one-loop low-energy effective action, explicitly, the potential and the kinetic term. Section 4 is a Summary where we discuss our results and perspectives.

\section{The $z=3$ Lifshitz four-fermion model}

Let us consider the $z=3$ Lifshitz four-fermion model, whose Lagrangian is given by
\begin{equation}\label{L0}
{\cal L}_0 = \bar\psi (i\slashed{\partial}_0+(i\slashed{\partial}_i)^3-m^3)\psi - \frac{g_t}{2N} (\bar\psi\lambda_0\gamma_0\psi)^2 - \frac{g_s}{2N} (\bar\psi  \¡\stackrel{\leftrightarrow}{\partial}_i\stackrel{\leftrightarrow}{\partial}_j\¡ \gamma^{ijk} \psi)^2,
\end{equation}
where $\slashed{\partial}_0=\partial_0\gamma^0$, $\slashed{\partial}_i=\partial_i\gamma^i$, $\stackrel{\leftrightarrow}{\partial}_i\,=\frac12 (\stackrel{\rightarrow}{\partial}_i\!-\!\stackrel{\leftarrow}{\partial}_i)$, and $\gamma^{ijk}=\lambda_1\gamma^i\gamma^j\gamma^k+\lambda_2\gamma^i\gamma^k\gamma^j+\lambda_3\gamma^k\gamma^i\gamma^j$, with $\lambda_i$ being constants, which can be either all together real or all together imaginary, to ensure reality of the Lagrangian. For a more convenient description of the dynamics we can proceed  in a manner similar to \cite{NJL,Thirring}, that is, we introduce the auxiliary vector fields, $A_0$ and $A_i$, allowing to rewrite (\ref{L0}) as follows:
\begin{eqnarray}\label{L02}
\mathcal{L}&=&\mathcal{L}_0+\frac{1}{2g_t}\left(A_0-\frac{g_t}{\sqrt{N}}\bar{\psi}\lambda_0\gamma_0\psi\right)^2+\frac{1}{2g_s}\left(A_k+\frac{g_s}{\sqrt{N}}\bar\psi  \¡\stackrel{\leftrightarrow}{\partial}_i\stackrel{\leftrightarrow}{\partial}_j\¡ {\gamma^{ij}}_k \psi\right)^2 \\
&=&\frac{1}{2g_t}A_0^2+\frac{1}{2g_s}A_k^2+\bar{\psi}(i\slashed{\partial}_0-i\slashed{\partial}_i\partial_j^2-e\lambda_0\slashed{A}_0+eA_k \¡\stackrel{\leftrightarrow}{\partial}_i\stackrel{\leftrightarrow}{\partial}_j\¡ \gamma^{ijk}-m^3)\psi,  \nonumber
\end{eqnarray}
where $e=\frac{1}{\sqrt{N}}$, $A_0^2=A_0A^0$, and so on. As a result, we arrive at the theory where the Horava-Lifshitz-like gauge field arises as an emergent phenomenon \cite{Bjorken}.

After integration by parts, we get
\begin{eqnarray}\label{L02a}
\mathcal{L}&=&\frac{1}{2g_t}A_0^2+\frac{1}{2g_s}A_k^2+\bar{\psi}(i\slashed{\partial}_0-i\slashed{\partial}_i\partial_j^2-e\lambda_0\slashed{A}_0-e\Delta_{ij}A_k \gamma^{ijk}-m^3)\psi,
\end{eqnarray}
with
\begin{equation}
\Delta_{ij}A_k = -\frac14 (\partial_i\partial_j A_k)-\frac12 (\partial_i A_k)\partial_j-\frac12 (\partial_j A_k)\partial_i-A_k\partial_i\partial_j.
\end{equation}

The corresponding generating functional is given by
\begin{eqnarray}
Z(\bar \eta,\,\eta) &=& \int DA_\mu D\psi D\bar\psi\, e^{i\int d^4x({\cal L}+\bar\eta\psi+\bar\psi\eta)}\nonumber\\
&=& \int DA_\mu\, e^{i\int d^4x \left(\frac{1}{2g_t}A_0^2+\frac{1}{2g_s}A_i^2\right)} \int D\psi D\bar\psi\, e^{i\int d^4x(\bar\psi S^{-1}\psi+\bar\eta\psi+\bar\psi\eta)},
\end{eqnarray}
where $S^{-1}=i\slashed{\partial}_0-i\slashed{\partial}_i\partial_j^2-e\lambda_0\slashed{A}_0-e\Delta_{ij}A_k \gamma^{ijk}-m^3$ is the operator describing the quadratic action.  To integrate over the fermion fields, we make the shift $\psi\rightarrow \psi-S\eta$ and $\bar{\psi}\rightarrow \bar{\psi}-\bar{\eta}S$, so that we arrive at the transformation $\bar\psi S^{-1}\psi+\bar\eta\psi+\bar\psi\eta \rightarrow \bar\psi S^{-1}\psi-\bar\eta S \eta$. {As a result, we obtain
\begin{eqnarray}
Z(\bar \eta,\,\eta) &=& \int DA_\mu\, e^{i\int d^4x \left(\frac{1}{2g_t}A_0^2+\frac{1}{2g_s}A_i^2\right)} \int D\psi D\bar\psi\, e^{i\int d^4x(\bar\psi S^{-1}\psi-\bar\eta S \eta)}.
\end{eqnarray}
Finally, integrating over fermions, we find the result for the generating functional
\begin{equation}\label{GF}
Z(\bar \eta,\,\eta) = \int DA_\mu \exp\left(iS_\mathrm{eff}[A] - i\int d^4x\, \bar\eta\, S\, \eta \right),
\end{equation}
where the one-loop effective action of the vector field is given by
\begin{equation}\label{Seff}
S_\mathrm{eff}[A] = \int d^Dx\, \left(\frac{1}{2g_t}A_0^2+\frac{1}{2g_s}A_i^2\right) -i \mathrm{Tr} \ln(\slashed{p}_0+\slashed{p}_ip_j^2-e\lambda_0\slashed{A}_0-e\Delta_{ij}(p)A_k \gamma^{ijk}-m^3),
\end{equation}
with
\begin{equation}
\Delta_{ij}(p)A_k = -\frac14 (\partial_i\partial_j A_k)+\frac i2 p_i(\partial_j A_k)+\frac i2 p_j(\partial_i A_k)+p_ip_jA_k
\end{equation}
and the space-time is $D$-dimensional.

Our task in the next section will consist in calculating the fermionic determinant in (\ref{Seff}), so that the lower terms of the derivative expansion of the one-loop effective action will be obtained explicitly.

\section{One-loop effective action}

Let us study the effective action. For this, we can rewrite (\ref{Seff}) as
\begin{equation}\label{acaoefetiva}
S_\mathrm{eff}[A] = \int d^Dx\, \left(\frac{1}{2g_t}A_0^2+\frac{1}{2g_s}A_i^2\right)+S_{\mathrm{eff}}^{(n)}[A],
\end{equation}
with
\begin{eqnarray}\label{acao_l}
S_{\mathrm{eff}}^{(n)}[A]&=&iN\mathrm{Tr}\sum_{n=1}^{\infty}\frac{1}{n}\left[S(p)e(\lambda_0\slashed{A}_0+\Delta_{ij}(p)A_k \gamma^{ijk})\right]^n,
\end{eqnarray}
where we have disregarded $-iN\mathrm{Tr}\ln(\slashed{p}_0+\slashed{p}_ip_j^2-m^3)$, since it is field independent. For $n=1$ and $n=3$, trivially, $S^{(3)}_\mathrm{eff}[A]$ and $S^{(1)}_\mathrm{eff}[A]$ vanish, since the trace of the corresponding product of odd number of Dirac matrices is zero.

Let us focus our attention on contributions with $n=2$ and $n=4$, whose analysis is sufficient for the generation of the kinetic and the lower-order potential terms. For $n=2$, we have
\begin{eqnarray}
S^{(2)}_\mathrm{eff}[A] &=& \frac {iN}2\mathrm{Tr}\,S(p)e(\lambda_0\slashed{A}_0+\Delta_{ij}(p)A_k \gamma^{ijk})S(p)e(\lambda_0\slashed{A}_0+\Delta_{ij}(p)A_k \gamma^{ijk}) \nonumber\\
&=& \frac{i}{2}\int d^Dx \Pi^{\mu\nu}A_\mu A_\nu,
\end{eqnarray}
where 
\begin{equation}
\Pi^{\mu\nu} = \mathrm{tr} \int \frac{d^Dp}{(2\pi)^D} S(p) \Gamma^\mu(p) S(p-i\partial) \Gamma^\nu(p-i\partial),
\end{equation}
with $\Gamma^\mu(p)=(\lambda_0\gamma^0,\Delta_{ij}(p)\gamma^{ijk})$. Therefore, since $S^{(2)}_\mathrm{eff}=\int d^4x {\cal L}^{(2)}_\mathrm{eff}$, we obtain the following low-energy effective Lagrangian:
\begin{eqnarray}\label{Leff2}
{\cal L}^{(2)}_\mathrm{eff} &=& \frac{1}{2}\alpha_1 A_iA^i -\frac{1}{2}(\alpha_2\partial_0A_i\partial^0A^i-\alpha_3\partial_0A_i\partial^iA^0-\alpha_3\partial_iA_0\partial^0A^i+\alpha_4\partial_iA_0\partial^iA^0) \\
&&-\frac{1}{2}(\alpha_5\partial_iA_j\partial^iA^j-\alpha_6\partial_iA_j\partial^jA^i) + \frac{1}2\mathrm{tr}\gamma^0\gamma^i\gamma^j(\alpha_7A_0\partial_i A_j+\alpha_7A_i\partial_j A_0-\alpha_8A_i\partial_0 A_j), \nonumber
\end{eqnarray}
where
\begin{subequations}\label{alphas}
\begin{eqnarray}
\alpha_1 &=&-\frac{2^{-d-1} \pi ^{-\frac{d}{2}-\frac{1}{2}} m^{d+1} \Gamma \left(\frac{1}{6} (-d-1)\right) \Gamma \left(\frac{d+4}{6}\right)}{9 \Gamma \left(\frac{d+2}{2}\right)} \nonumber\\
&&\times\left((d-2) \lambda _2^2-2 (2 d-1) \left(\lambda _1+\lambda _3\right) \lambda _2+(d-2) \left(\lambda _1+\lambda _3\right){}^2\right) \mathrm{tr}{\bf1}, \\
\alpha_2 &=& \frac{2^{-d-1} \pi ^{-\frac{d}{2}-\frac{1}{2}} m^{d-5} \Gamma \left(\frac{5}{6}-\frac{d}{6}\right) \Gamma \left(\frac{d+4}{6}\right)}{27 \Gamma \left(\frac{d+2}{2}\right)} \nonumber\\
&&\times\left((2 d-1) \lambda _2^2-(5 d-7) \left(\lambda _1+\lambda _3\right) \lambda _2+(2 d-1) \left(\lambda _1+\lambda _3\right){}^2\right) \mathrm{tr}{\bf1}, \\
\alpha_3 &=& -\frac{2^{-d-2} \pi ^{-\frac{d}{2}-\frac{1}{2}} m^{d-5} \Gamma \left(\frac{5}{6}-\frac{d}{6}\right) \Gamma \left(\frac{d+4}{6}\right)}{9 \Gamma \left(\frac{d+2}{2}\right)} \lambda _0 \left(2 (d-2) \lambda _2-(d+1) \left(\lambda _1+\lambda _3\right)\right) \mathrm{tr}{\bf1}, \\
\alpha_4 &=& \frac{2^{-d} \pi ^{-\frac{d}{2}-\frac{1}{2}} m^{d-5} \Gamma \left(\frac{5}{6}-\frac{d}{6}\right) \Gamma \left(\frac{d+4}{6}\right)}{3 d \Gamma \left(\frac{d}{2}\right)} \lambda _0^2\mathrm{tr}{\bf1},
\end{eqnarray}
and
\begin{eqnarray}
\alpha_5 &=& \frac{2^{-d-3} \pi ^{-\frac{d}{2}-\frac{1}{2}} m^{d-1} \Gamma \left(\frac{1}{6}-\frac{d}{6}\right) \Gamma \left(\frac{d+8}{6}\right)}{9 \Gamma \left(\frac{d}{2}+2\right)} \left((d (d+4)-2) \lambda _1^2+2 \lambda _1 \left((d (d+4)-2) \lambda _3 \right.\right. \\ 
&&\left.\left.-(d-1) (2 d+7) \lambda _2\right)+(d (d+4)-8) \lambda _2^2+(d (d+4)-2) \lambda _3^2-2 (d-1) (2 d+7) \lambda _2 \lambda _3\right) \mathrm{tr}{\bf1}, \nonumber\\
\alpha_6 &=& \frac{2^{-d-3} \pi ^{-\frac{d}{2}-\frac{1}{2}} m^{d-1} \Gamma \left(\frac{1}{6}-\frac{d}{6}\right) \Gamma \left(\frac{d+8}{6}\right)}{9 \Gamma \left(\frac{d+4}{2}\right)} \left(-4 (d-1)^2 \lambda _2 \lambda _3+((d-2) d+4) \lambda _1^2 \right. \nonumber\\
&&\left.+((d-8) d+4) \lambda _2^2+((d-2) d+4) \lambda _3^2+2 \lambda _1 \left(((d-2) d+4) \lambda _3-2 (d-1)^2 \lambda _2\right)\right) \mathrm{tr}{\bf1}, \\
\alpha_7 &=& \frac{2^{-d-2} \pi ^{-\frac{d}{2}-\frac{1}{2}} m^{d-2} \Gamma \left(\frac{5}{6}-\frac{d}{6}\right) \Gamma \left(\frac{d+4}{6}\right)}{3 \Gamma \left(\frac{d}{2}+1\right)} \lambda_0 \left(d \lambda _2-(d+2) \left(\lambda _1+\lambda _3\right)\right), \\
\alpha_8 &=& \frac{2^{-d-2} \pi ^{-\frac{d}{2}-\frac{1}{2}} m^{d-2} \Gamma \left(\frac{5}{6}-\frac{d}{6}\right) \Gamma \left(\frac{d+4}{6}\right)}{3 \Gamma \left(\frac{d}{2}+1\right)} \left(\lambda _1-\lambda _2+\lambda _3\right) \left(d \lambda _1-(d-4) \lambda _2+d \lambda _3\right),
\end{eqnarray}
\end{subequations}
where we have considered $D=d+1$ and ${\bf 1}$ is the unit matrix of the corresponding dimension, i.e., the $2^{D/2}\times2^{D/2}$ matrix.

Requiring the gauge invariance of our result, we must impose the equality $\alpha_2=\alpha_3=\alpha_4$ and $\alpha_5=\alpha_6$. Thus, we get $\lambda_0=\lambda_2$ and $\lambda_3=2\lambda_2-\lambda_1$, which implies $\alpha_7=\alpha_8$, that is the relation necessary to ensure the gauge invariant Chern-Simons term, so that we obtain
\begin{eqnarray}\label{Leff2a}
{\cal L}^{(2)}_\mathrm{eff} &=& \frac{1}{2}\alpha_1 A_iA^i -\frac{1}{2}\alpha_3 F_{0i}F^{0i} -\frac{1}{4}\alpha_5F_{ij}F^{ij} +\frac{i}{2}\alpha_7\mathrm{tr}\slashed{A}\slashed{\partial}\slashed{A},
\end{eqnarray}
where
\begin{subequations}\label{alphas2}
\begin{eqnarray}
\alpha_1 &=& \frac{2^{-d} (d+2) \pi ^{\frac{1}{2} (-d-1)} \lambda _2^2 m^{d+1} \Gamma \left(\frac{1}{6} (-d-1)\right) \Gamma \left(\frac{d+4}{6}\right)}{3 d \Gamma \left(\frac{d}{2}\right)}, \\
\alpha_3 &=& \frac{2^{-d} \pi ^{\frac{1}{2} (-d-1)} \lambda _2^2 m^{d-5} \Gamma \left(\frac{5}{6}-\frac{d}{6}\right) \Gamma \left(\frac{d+4}{6}\right)}{3 d \Gamma \left(\frac{d}{2}\right)}, \\
\alpha_5 &=& -\frac{2^{-d-2} (d-2) \pi ^{\frac{1}{2} (-d-1)} \lambda _2^2 m^{d-1} \Gamma \left(\frac{1-d}{6}\right) \Gamma \left(\frac{d+8}{6}\right)}{3 \Gamma \left(\frac{d}{2}+1\right)}, \\
\alpha_7 &=& \frac{2^{-d} \pi ^{\frac{1}{2} (-d-1)} \lambda _2^2 m^{d-2} \Gamma \left(\frac{5}{6}-\frac{d}{6}\right) \Gamma \left(\frac{d+10}{6}\right)}{d \Gamma \left(\frac{d}{2}\right)}.
\end{eqnarray}
\end{subequations}
Note that now all coefficients $\alpha_i$ are written in terms of $\lambda_2^2$.

Let us finally consider $n=4$, by writing the effective action as
\begin{eqnarray}
S^{(4)}_\mathrm{eff}[A] &=& \frac{iN}4 \mathrm{Tr}\,S(p)e(\lambda_0\slashed{A}_0+\Delta_{ij}(p)A_k \gamma^{ijk})S(p)e(\lambda_0\slashed{A}_0+\Delta_{ij}(p)A_k \gamma^{ijk}) \nonumber\\
&&\times S(p)e(\lambda_0\slashed{A}_0+\Delta_{ij}(p)A_k \gamma^{ijk})S(p)e(\lambda_0\slashed{A}_0+\Delta_{ij}(p)A_k \gamma^{ijk}) \nonumber\\
&=& \frac{ie^2}{4}\int d^4 \Pi^{\kappa\lambda\mu\nu}A_\kappa A_\lambda A_\mu A_\nu,
\end{eqnarray}
where
\begin{equation}
\Pi^{\kappa\lambda\mu\nu} = \mathrm{tr} \int \frac{d^4p}{(2\pi)^4} S(p) \Gamma^\kappa(p)S(p)\Gamma^\lambda(p)\Gamma^\mu(p)S(p)\Gamma^\nu(p)+{\cal O}(\partial^4).
\end{equation}
Then, we obtain
\begin{equation}\label{Leff4}
{\cal L}^{(4)}_\mathrm{eff} =  -\frac{e^2}{4}\beta A_iA^i A_jA^j,
\end{equation}
with
\begin{eqnarray}\label{beta}
\beta &=& \frac{2^{-d-1} (d-1) \pi ^{\frac{1}{2} (-d-1)} m^{d-1} \Gamma \left(\frac{1}{6}-\frac{d}{6}\right) \Gamma \left(\frac{d+8}{6}\right)}{27 \Gamma \left(\frac{d+4}{2}\right)} \left((d-4) \lambda _2^4-4 (2 d+1) \left(\lambda _1+\lambda _3\right) \lambda _2^3 \right. \nonumber\\
&&\left.+18 (d+2) \left(\lambda _1+\lambda _3\right){}^2 \lambda _2^2-4 (2 d+1) \left(\lambda _1+\lambda _3\right){}^3 \lambda _2+(d-4) \left(\lambda _1+\lambda _3\right){}^4\right) \mathrm{tr}{\bf1},
\end{eqnarray}
so that, when $\lambda_3=2\lambda_2-\lambda_1$, we get
\begin{equation}
\beta = -\frac{2^{-d} (d+4) \pi ^{\frac{1}{2} (-d-1)} \lambda _2^4 m^{d-1} \Gamma \left(\frac{7}{6}-\frac{d}{6}\right) \Gamma \left(\frac{d+8}{6}\right)}{\Gamma \left(\frac{d}{2}+2\right)}.
\end{equation}
Observe that surprisingly $\beta$ is only depending on $\lambda_2$, as well as the coefficients (\ref{alphas2}), so that the gauge invariance is satisfied.

Therefore, considering Eq.~(\ref{acaoefetiva}), the effective potential is
\begin{equation}\label{Veff}
V_\mathrm{eff}=-\frac{1}{2g_t}A_0^2-\frac{1}{2g_s}A_i^2-\frac{1}{2}\alpha_1 A_i^2+\frac{e^2}{4}\beta A_i^4.
\end{equation}
Thus, we have the gap equations
\begin{eqnarray}
\label{gap}
\frac{d V_{eff}}{dA_0}\bigg|_{A_\mu=a_\mu} = -\frac1{g_t} a^0 = 0 \\
\frac{d V_{eff}}{dA_i}\bigg|_{A_\mu=a_\mu} = \left(-\frac1{g_s}-\alpha_1+e^2\beta a_j^2 \right) a^i = 0,
\end{eqnarray}
i.e., for $a_0\neq0$ and $a_i\neq0$, we obtain the conditions $g_t\to\infty$ and $\frac1{g_s}=-\alpha_1+e^2\beta a_j^2$. With this, we can rewrite (\ref{Veff}) as 
\begin{equation}
V_\mathrm{eff}=\frac{e^2}{4}\beta\left(A_i^2-a_i^2\right)^2 -\frac{e^2}{4}\beta a_i^4.
\end{equation}
Then, from (\ref{acaoefetiva}), we get the effective Lagrangian 
\begin{eqnarray}\label{Leff}
{\cal L}_\mathrm{eff} &=& -\frac{1}{2}\alpha_3 F_{0i}F^{0i} -\frac{1}{4}\alpha_5F_{ij}F^{ij} +\frac{i}{2}\alpha_7\mathrm{tr}\slashed{A}\slashed{\partial}\slashed{A} -\frac{e^2}{4}\beta\left(A_i^2-a_i^2\right)^2,
\end{eqnarray}
where we have dropped away the constant term $-\frac{e^2}{4}\beta a_i^4$.

For $D=3$ (or $d=2$), $\alpha_5=0$ and $\beta<0$, so that we obtain
\begin{eqnarray}
{\cal L}_\mathrm{eff} &=& -\frac{1}{2}\alpha_3 F_{0i}F^{0i} -\frac{1}{2}\alpha_7\epsilon^{\lambda\mu\nu}A_\lambda\partial_\mu A_\nu -\frac{e^2}{4}\beta\left(A_i^2-a_i^2\right)^2,
\end{eqnarray}
i.e., in this case the Chern-Simons term is Lorentz invariant, but there is no Lorentz-invariant Maxwell term. As $\beta<0$, we cannot have potential with spontaneous symmetry breaking in this space-time dimension. Besides this, the $D=3$ result is not satisfactory for us since we have no Maxwell term arisen. However, we see that in this case the Maxwell-like term is described by the electric field only and in the limit $N\to\infty$, that is, $e^2\to 0$, the potential term vanishes. Thus, we arrive at a particular Maxwell-Chern-Simons-like theory.

Let us now suggest the dimension $D$ to be arbitrary and even, so, the Chern-Simons term proportional to the trace of the product of three Dirac matrices will vanish. Then, we rewrite Eq.~(\ref{Leff}) as follows:
\begin{eqnarray}\label{Leff2b}
{\cal L}_\mathrm{eff} &=& -\frac{1}{2}\alpha_3 F_{0i}F^{0i} -\frac{1}{4}\alpha_5F_{ij}F^{ij} -\frac{e^2}{4}\beta\left(A_i^2-a_i^2\right)^2,
\end{eqnarray}
which, for $D=4$, we have $\beta<0$, i.e., there is no spontaneous symmetry breaking potential as well. The interesting case is $D=10$, where we have either $\alpha_{3,5}<0$ and $\beta>0$ (for $\lambda_2=1$) or $\alpha_{3,5}>0$ and $\beta>0$ (for $\lambda_2=i$), which means that the Lagrangian (\ref{Leff2}) now has a positively defined potential and Lorentz violating kinetic term, when we choose $\lambda_2=i$. We note that in both these cases $\alpha_3$ and $\alpha_5$ have the same sign which allows for a rescaling of fields and derivatives in order to get the Maxwell-like term $\alpha F_{\mu\nu}F^{\mu\nu}$, with the overall sign of this term being positive for $\lambda_2=1$, and negative, matching thus the standard form, for $\lambda_2=i$.

In order to rewrite the kinetic term of (\ref{Leff2}) in the Maxwell-like form, we carry out the rescaling of fields and constant parameters just as in \cite{ourHLED}. Explicitly, rewriting (\ref{alphas2}) and (\ref{beta}) as $\alpha_3=\lambda_2^{2}m^{d-5}\tilde\alpha_3$, $\alpha_5=\lambda_2^{2}m^{d-1}\tilde\alpha_5$, $\alpha_7=\lambda_2^{2}m^{d-2}\tilde\alpha_7$, and $\beta=\lambda_2^{4}m^{d-1}\tilde\beta$, and considering the rescaling 
\begin{subequations}\label{reescal}
\begin{eqnarray}
A_0 &\to& \frac{m^2\tilde\alpha_5^{1/4}}{\tilde\alpha_3^{1/2}}A_0, \\
A_i &\to& \frac{1}{\tilde\alpha_5^{1/4}}A_i, \\
\partial_0 &\to& \frac{m^2\tilde\alpha_5^{1/4}}{\tilde\alpha_3^{1/2}}\partial_0, \\
\partial_i &\to& \frac{1}{\tilde\alpha_5^{1/4}}\partial_i,
\end{eqnarray}
\end{subequations}
the effective Lagrangian (\ref{Leff2}) can be rewritten as
\begin{eqnarray}
\label{actresc}
{\cal L}_\mathrm{eff} &=& -\frac{1}{4}m^{d-1}F_{\mu\nu}F^{\mu\nu} -\frac{e^2\lambda_2^2\tilde\beta}{4\tilde\alpha_5}m^{d-1}\left(A_i^2-a_i^2\right)^2.
\end{eqnarray}
It is straightforward to see that the factor 
\begin{equation}
\frac{e^2\lambda_2^2\tilde\beta}{4\tilde\alpha_5} = -\frac{(d-1) (d+4) e^2\lambda _2^2}{(d-2)(d+2)}
\end{equation}
 is positive for all dimensions, including $D=d+1$ odd, as $\lambda_2=i$. Actually, we found that, after the rescaling (\ref{reescal}), the potential is always positively defined, in contrast of the situation before (\ref{Leff2}), where its positivity only occurs for $D=10$.
 
We note nevertheless that the case $\lambda_2=i$, necessary to guarantee the positivity of our effective potential, does not break the Hermiticity of the initial Lagrangian (\ref{L0}) since actually it implies that the fields $A_i,A_0$ are purely imaginary, which, in its part, implies that $A^2_i<0$. Therefore, to have minima, one should have $a^2_i<0$ in (\ref{actresc}). Let us verify the consistency of this requirement. As we have noted after (\ref{gap}), the $a_i$, that is, the value of our vector field  corresponding to the minimum of the potential, satisfies the equation $a^2_j=\frac{1}{e^2\beta}(\frac{1}{g_s}+\alpha_1)$. Clearly, since $A_j$ is imaginary, the $a_j$ should be imaginary as well, i.e. $a^2_j<0$. Straightforward checking shows that for $g_s>0$, the restriction $a^2_j<0$ can be consistent with the equation above  for $d=3$, where $\beta<0$, for $g_s$ small enough. For $d=9$, where $\beta>0$, vice versa, this restriction is valid if $g_s<0$ also with its absolute value small enough.
Nevertheless, we note that there is no fundamental restriction on the sign of $g_s$ from the basic reasons, hence our calculations are consistent.

\section{Summary}

We considered the theory where a dynamical vector field arises as a Lagrange multiplier in the $z=3$ Lifshitz four-fermion model. For this field, we introduced the one-loop effective action in terms of the fermionic determinant and explicitly found the lowest terms in its derivative expansion in various spatial dimensions. We found that in three dimensions, the Lorentz-invariant Chern-Simons term arises, which is an advantage of our scheme in comparison with \cite{ourGN}, where the Chern-Simons-like one-derivative term displayed neither Lorentz nor gauge symmetry. However, a Lorentz-invariant Maxwell term is not generated at $D=3$, i.e., only $ -\frac{1}{2}\alpha_3 F_{0i}F^{0i}$ is present. In other dimensions, nevertheless, the trace of the product of three Dirac matrices accompanying the one-derivative term clearly vanishes.

The results we achieved in this paper represent themselves as a further development of the concept of the emergent dynamics \cite{Bjorken}, but the advantage of our result consists in the fact that, while in \cite{Bjorken} the initial theory was Lorentz invariant itself, we presented a mechanism allowing to obtain Lorentz-invariant kinetic term on the base of non-Lorentz-invariant theory, which allows treating the Lorentz symmetry as an emergent phenomenon without its suggestion from the very beginning. Therefore, we provided a more realistic mechanism for a dynamical generation of the electromagnetic field.

The most interesting result of our paper is the possibility to arrive at the positively defined bumblebee-like potential in any space-time dimension, see Eq.~(\ref{actresc}). We note that higher space-time dimensions were not considered earlier within the context of Horava-Lifshitz-like theories, to the best of our knowledge. Thus, in principle, our study opens new horizons for application of the Horava-Lifshitz methodology within the context of the extra dimensions problem.


{\bf Acknowledgements.} This work was partially supported by Conselho
Nacional de Desenvolvimento Cient\'{\i}fico e Tecnol\'{o}gico (CNPq).  The work by A. Yu. P. has been partially supported by the CNPq project 303783/2015-0.

\end{document}